\documentclass[ES]{rcftex}

\title{Optimizaci\'on de un modelo para los planos $CuO$ en el $La_2CuO_4$ }

\titleEN{Optimization of a model for the  $CuO$ planes in $La_2CuO_4$} 

\author{Y. Vielza$^{a}$ y A. Cabo Montes de Oca$^{b,\dag}$ \medskip}

\affiliations{
\aff     Facultad de F\'isica, UH, Colina Universitaria, Vedado, La Habana, Cuba \,\,  yvielzadelacruz@gmail.com
\aff    Departamento de F\'{\i}sica Te\'orica,
Instituto de Cibern\'{e}tica, Matem\'{a}tica y F\'{\i}sica, Calle E,
No. 309, Vedado, La Habana, Cuba.  \,\, cabo@icimaf.cu
}

\abstractES{\noindent Se extiende un estudio previo donde se consider\'o el efecto del dopaje con
huecos en un modelo simple de las capas $CuO_{2}$ del $La_{2}CuO_{4}$. Se reajustan los par\'ametros con el objetivo de fijar los valores del gap del material en $2\,\, eV$ y su constante diel\'ectrica cercana a $21$. Se reobtienen   las indicaciones de una  transici\'on de fase ``\,escondida'' dentro del ``\,Domo"  superconductor. La transici\'on es de segundo orden y est\'a asociada a la coincidencia energ\'etica de un estado b\'asico  aislante (AFA) con un estado excitado paramagn\'etico que muestra un Pseudogap(PPG) en un punto cr\'itico de concentraci\'on de huecos cercano a $x_{c}=0.2$.  Se presenta la evoluci\'on con el dopaje de las bandas y las superficies de Fermi en las fases AFA y PPG. En la zona de bajo dopaje los huecos comienzan a ocupar los estados situados en la frontera de la zona de Brillouin, que en el estado AFA son los que cargan el antiferromagnetismo m\'as intenso. Alrededor del dopaje cr\'itico los resultados muestran que ambas fases tienden  a coincidir en sus superficies de Fermi y los espectros de energ\'ia de los estados
ocupados.
}

\abstractEN{The results of a previous  work, where  it was considered the effect of  hole doping on a simple model of the  $CuO_{2}$ planes in $La_{2}CuO_{4}$, are extended. The parameters are adjusted with the objective to fix the  known values of the  gap of $2\,\, eV$ for this material and its  dielectric constant of $21$. We find again indications of a ``\,hidden" phase transition beneath   the superconductor dome. The transition is a second order one, and is associated with an energetic coincidence of a ground insulator state (AFA)
with an excited paramagnetic state showing a pseudogap (PPG),  at  a critical point of the hole concentration around  $x_{c}=0.2$. The evolution as a  function of doping of the band structures and the Fermi  surface
of the system in the phases AFA and PPG, is shown. In the zone of low doping,  the holes begin to occupy the
states located at the mid of the sides of the Brillouin zone, that in the
AFA states have the  strongest antiferromagnetic character. Around the critical doping the results show that in both phases the Fermi surfaces and the energy spectrum of the filled  electronic states tend to coincide. }

\keyW{ Cuprate superconductors 74.72.-h, Strongly correlated electron
systems 71.27.+a, Metal–insulator transitions  71.30.+h, La-based cuprates 74.72.Dn,
 Theories and models of many electron
systems 71.10.-w}

\begin{document}

\maketitle

\section{Introducci\'on}

En 1986 Georg Bednorz y Alex M\"{u}ller, al investigar compuestos basados en
el \'{o}xido de cobre, descubren los superconductores de alta temperatura
cr\'{\i}tica (HTSC por sus siglas en ingl\'{e}s) \cite{key-14}, generando un enorme
inter\'{e}s en este tipo de materiales conocidos como \textquotedblleft
cupratos\textquotedblright. \ Estos sistemas cuentan con una estructura
cristalina en la que se observan capas de \'{o}xidos de cobre que controlan el
comportamiento del material ante el paso de la corriente el\'{e}ctrica. En el
estado normal la conducci\'{o}n el\'{e}ctrica en estos planos es
aproximadamente cien veces mayor que en la direcci\'{o}n perpendicular. Por
esta raz\'{o}n se dice que, en cuanto a la conducci\'{o}n el\'{e}ctrica, los
cupratos son sistemas cuasi-bidimensionales \cite{key-15}. Estos materiales tienen
diferencias notables respecto a los superconductores que hab\'{\i}an sido
encontrados no s\'{o}lo por su  alta temperatura cr\'{\i}tica, que no es explicable
por la famosa teor\'{\i}a BCS, sino tambi\'{e}n debido a sus no convencionales
propiedades f\'isicas  en la fase normal.

En contra de lo esperado a priori, los cupratos son aislantes de Mott y los
electrones localizados se ordenan de forma antiferromagn\'{e}tica. Un aislante
de Mott es un sistema electr\'{o}nico que se encuentra en una fase en la cual
hay un gap en el espectro de energ\'{\i}as de una part\'{\i}cula y este gap
est\'{a} generado por las fuertes correlaciones electr\'{o}nicas y no por las
caracter\'{\i}sticas de la red como en los aislantes usuales. El paso de la
corriente el\'{e}ctrica en este tipo de materiales se inhibe para evitar que
haya dos electrones en el mismo \'{a}tomo ya que debido a la fuerte
repulsi\'{o}n esto costar\'{\i}a mucha energ\'{\i}a. Por su parte la fuerte
tendencia de los cupratos a tener estados electr\'{o}nicos ordenados se
evidencia de las famosas fases tipo $nem\acute{a}ticas$ o $de\,\,stripes$ las
cuales rompen alguna simetr\'{\i}a espacial del sistema. Estas fases han sido
intensamente estudiadas en superconductores de alta temperatura y actualmente
pueden encontrarse en la literatura de distintos trabajos de resumen  acerca
del tema \cite{key-16,key-17}. La relaci\'{o}n entre estos estados ordenados y los
mecanismos que generan la superconductividad de alta temperatura son en la
actualidad uno de los temas de mayor  inter\'{e}s en la F\'{\i}sica de la
Materia Condensada.

As\'{\i}, a pesar de la investigaci\'{o}n intensiva y de muchas ideas
prometedoras que buscan explicar la existencia de la superconductividad no
convencional, aun no se ha logrado un consenso respecto a la tesis m\'{a}s
apropiada. Una de las teor\'{\i}as que parece tener la base necesaria para
alcanzar este fin est\'{a} basada  en el proceso de dopar con huecos un
aislante de Mott,  y en ella  la superconductividad se genera directamente de
la fuerte interacci\'{o}n repulsiva de los electrones.

Dentro de la amplia familia de cupratos se encuentra el $La_{2}CuO_{4}$ quien
figura como uno de los compuestos m\'{a}s estudiados experimentalmente. Su
simple estructura cristalina y regulada concentraci\'{o}n de huecos sobre los
planos bidimensionales $CuO_{2}$, en un amplio r\'{e}gimen de dopaje, sugieren
que una posible condensaci\'{o}n de pares de huecos enlazados den lugar a
propiedades de transporte superconductoras guiadas sobre las capas $Cu$-$O$.

Interesante resulta la variedad de fases de este material en la regi\'{o}n de
temperaturas cercanas al cero absoluto donde, en la medida que aumentamos la
concentraci\'{o}n de huecos, un estado AFA,  existente a bajo dopaje,
evoluciona  hacia un estado superconductor y luego a un metal normal. No
obstante, entre los aspectos m\'{a}s enigm\'{a}ticos del diagrama de fases
destaca una posible transici\'{o}n de fase cu\'{a}ntica dentro del Domo
superconductor que se estima ocurre a cero temperatura en un punto cr\'{\i}tico de concentraci\'{o}n de
huecos \cite{key-38}.  Resulta as\'i necesario esclarecer los or\'{\i}genes de la conducci\'{o}n en estos materiales y su evoluci\'{o}n en la medida que se dopa el compuesto con vistas a descifrar  la forma
compleja que adopta su estructura. En particular nosotros estimamos  que la
existencia de estados ligados de huecos preformados en la fase AF aislante de
Mott y su posterior condensaci\'{o}n de Bose-Einstein muestran una ruta
prometedora hacia la superconductividad.

Haciendo uso de un modelo de una banda resuelto en aproximaci\'{o}n a HF, en las
referencias \cite{key-1,key-2,key-3} fue posible  predecir la existencia tanto del estado  aislante antiferromagn\'etico
como del estado de  pseudogap en este material. Posteriormente en  \cite{key-6,key-6b},
se introdujo el efecto de dopaje con huecos en dicho modelo,  lo cual permiti\'o describir
varias propiedades de mucho  inter\'es del $La_2CuO_4$ a $T=0\, K$. Sin embargo, cabe subrayar  que los resultados
experimentales fijan el gap del estado b\'{a}sico AFA del $La_2CuO_4$ a 2.0 eV (\cite{gap})y su constante
dil\'{e}ctrica $\epsilon$ a un valor cercano a 20 (\cite{dielect}). Los par\'ametros utilizados en \cite{key-6}, aunque fueron semejantes
 no coincidieron con esos valores (1.3 eV de gap y $\epsilon$ aproximadamente igual a $10$).
 De esa manera, la motivaci\'on central del presente trabajo  la constituy\'{o} fijar con m\'as
precisi\'{o}n estas dos propiedades  con vistas a establecer m\'{a}s
apropiadamente dichos  par\'{a}metros.
En consecuencia, una vez optimizada las bases del modelo, tambi\'en hicimos una correcci\'{o}n a los c\'{a}lculos que incluyen el dopaje con huecos del compuesto. Como se describir\'a m\'as adelante, en este proceso se prefij\'{o} nuevamente el ancho de la banda paramagn\'{e}tica obtenida de la soluci\'{o}n HF, al valor del ancho 3.8 eV de la \'{u}nica banda que cruza el nivel de Fermi en los c\'{a}lculos de bandas de Matheiss \cite{key-4}.

 Describamos a continuaci\'on c\'omo procede  la exposici\'on.
En la Secci\'on 2  hacemos  de inicio  una revisi\'{o}n del modelo de una banda introducido en las referencias \cite{key-1,key-2,key-3} y  de su soluci\'{o}n de campo medio. En esta secci\'on tambi\'en se realiza la fijaci\'on de los valores observados del gap (2 eV) y la constante diel\'ectrica ($\epsilon=21$) del material. La soluci\'on de campo medio  brinda entonces los estados aislante  y de pseudogap a semillenado (ausencia de dopaje).
Finalmente, en la Secci\'on 3 se investigan  los estados que predice el modelo para las capas $Cu$-$O$
en funci\'{o}n del dopaje con huecos. Se estudia la evoluci\'on de  la superficie
de Fermi  y de los estados uniparticulares HF en las fases AFA y PPG. Todo ello permite
 profundizar la argumentaci\'on dada en \cite{key-6,key-6b} acerca de la existencia de una  transici\'{o}n de fase cu\'{a}ntica dentro del Domo superconductor, tal como indican  los datos experimentales \cite{key-38}.

\section{ Modelo de Tight-Binding de los planos $CuO$}

\begin{figure}
\begin{centering}
\includegraphics[scale=0.4]{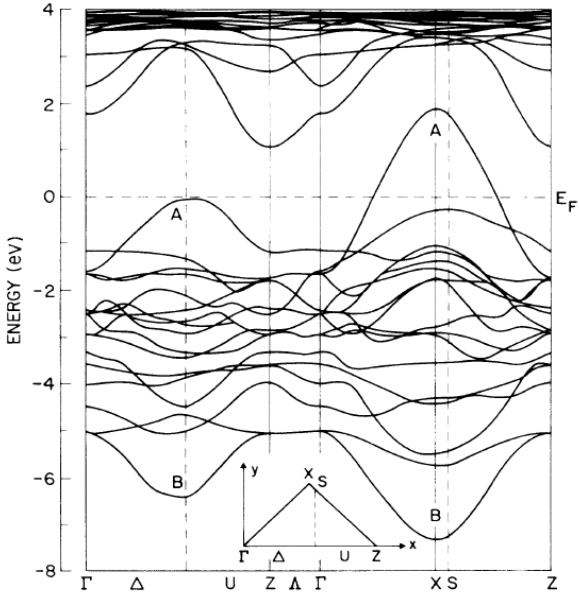}
\par\end{centering}
\caption{{\protect\small Estructura de bandas para el $La_{2}CuO_{4}$
calculada por Horsch y Stephan y cols. en 1993 y Matheiss y cols. en 1987. La
banda semillena, predice un comportamiento tight-binding en el plano
rec\'{\i}proco al }$CuO_{2}${\protect\small {} (direcci\'{o}n $\Gamma$-$X$).}}%
\label{matheiss}
\end{figure}

En esta secci\'on  describiremos el modelo simplificado del
plano cobre-ox\'{\i}geno electr\'{o}nico introducido en las referencias \cite{key-1,key-2,key-3}.  En la Figura \ref{matheiss}, se ilustra el diagrama de bandas asociado al $La_{2}%
CuO_{4}$ obtenido mediante t\'{e}cnicas de $LAPW$ (Linear Aumented Plane
Waves) {[}4{]}. Describamos a partir de ese diagrama la construcci\'on  del
modelo. Puede notarse que la \'{u}ltima banda ocupada
est\'{a} semillena,  lo cual predice un comportamiento met\'{a}lico del material
(hay una banda que atraviesa el nivel de Fermi en la Fig. \ref{matheiss}).
La forma de esta banda sugiere la validez de un
esquema de electrones fuertemente ligados (TB) para el gas de electrones que
la puebla. El electr\'{o}n menos ligado al compuesto $La_{2}CuO_{4}$, es aquel
que no se encuentra apareado en el $Cu^{2+}$, que a diferencia de los $O^{2-}$
en el plano, no tiene su \'{u}ltima capa $(3d)$ cerrada. Estos electrones en
un cuadro cualitativo, pueden estimarse como los que constituyen la \'{u}nica
banda del material que corta el nivel de Fermi en los c\'{a}lculos de la
referencia {[}4{]}. Por esto es razonable considerar que esos electrones
est\'{a}n fuertemente correlacionados a las celdas base $CuO_{2}$ y con
especial preferencia hacia los \'{a}tomos de $Cu$ correspondientes, asumida la
completitud de capas del $O^{2-}$. La anterior idea justifica tomar como la
red asociada al modelo TB que dar\'{\i}a lugar a la banda semillena de la
referencia {[}4{]}, como una red cuadrada de puntos coincidentes con los
sitios $Cu$ en el plano $CuO_{2}$.

La presencia de todos los dem\'{a}s electrones que llenan las otras bandas en
conjunto con las cargas nucleares que neutralizan la electr\'{o}nica, juega un
doble papel en el modelo. En primer lugar: como medio efectivo
polarizable al cual asociamos cierta permitividad diel\'{e}ctrica $\epsilon$
que apantalla el campo producido por cualquier carga puntual extra\~{n}a a
\'{e}l. En segundo lugar: por su distribuci\'{o}n espacial y magnitud, se
considera responsable  en garantizar con su acci\'{o}n el orden peri\'{o}dico
del s\'{o}lido. Esto se modela a partir de suponer que esos electrones y
cargas nucleares crean un potencial peri\'{o}dico confinante $W_{\gamma}$ en la red puntual.
El modelo se completa considerando las interacciones
internas del gas electr\'{o}nico que semillena la banda considerada y
adem\'{a}s su interacci\'{o}n con el excedente de cargas (jellium) que los
neutraliza $F_{b}$. A este lo modelamos  como una distribuci\'{o}n gaussiana de
cargas positivas alrededor de cada punto de la red, con radio
caracter\'{\i}stico $b$.

El Hamiltoniano del modelo tiene la forma%
\begin{equation}
\hat{\mathcal{H}}_{0}(\mathbf{x})=\frac{\mathbf{p}^{2}}{2m}+W_{\gamma
}(\mathbf{x})+F_{b}(\mathbf{x}), \label{eq:6}%
\end{equation}
donde se tiene
\begin{equation}
W_{\gamma}(\mathbf{x})=W_{\gamma}(\mathbf{x}+\mathbf{R}), \label{eq:7}%
\end{equation}
\begin{eqnarray}
F_{b}(\mathbf{x})=\frac{e^{2}}{4\pi\varepsilon_{0}\varepsilon}\sum
_{\mathbf{R}}\int d^{2}y\frac{\exp(-\frac{(\mathbf{y}-\mathbf{R})^{2}}{b^{2}%
})/\pi b^{2}}{\left|  \mathbf{x}-\mathbf{y}\right|  },
\label{eq:8}%
\end{eqnarray}
donde $ b\ll p$ y los vectores que describen las coordenadas de los \'{a}tomos de $Cu$ se
definen por
\begin{equation}
\mathbf{R}=\left\{
\begin{array}
[c]{c}%
(n_{x_{1}}p\,\mathbf{e}_{x_{1}}+n_{x_{2}}p\,\mathbf{e}_{x_{2}})\\
con\,\, n_{x_{1}},n_{x_{2}}\in {Z}
\end{array}
\right.  , \label{eq:9}%
\end{equation}
siendo $\mathbf{e}_{x_{1}}$ y $\mathbf{e}_{x_{2}}$ los versores que est\'{a}n
sobre las direcciones definidas por los vecinos m\'{a}s cercanos de esa red.
Se conoce que la distancia entre un \'{a}tomo de $Cu$ y su vecino m\'{a}s
pr\'{o}ximo es $p=3.82\,\text{\AA }$ \cite{key-26,key-37}. Por otra parte se considera que
la interacci\'{o}n entre un par de electrones del gas que semillena la banda
electr\'{o}nica en consideraci\'{o}n est\'{a} dada por el potencial de
Coulomb:
\begin{equation}
V=\frac{e^{2}}{4\pi\epsilon_{0}\epsilon}\frac{1}{\left|  \mathbf{x}%
-\mathbf{y}\right|  }, \label{eq:10}%
\end{equation}
la cual incluye una constante diel\'{e}ctrica que se asume determinada por la
respuesta electromagn\'etica  del gas de electrones y n\'{u}cleos, ya que estos
constituyen el medio en que se mueve el gas electr\'{o}nico de la banda
semillena en consideraci\'{o}n.
\begin{figure}
\begin{centering}
(a)\,\,\,\,\,\,\includegraphics[scale=0.7]{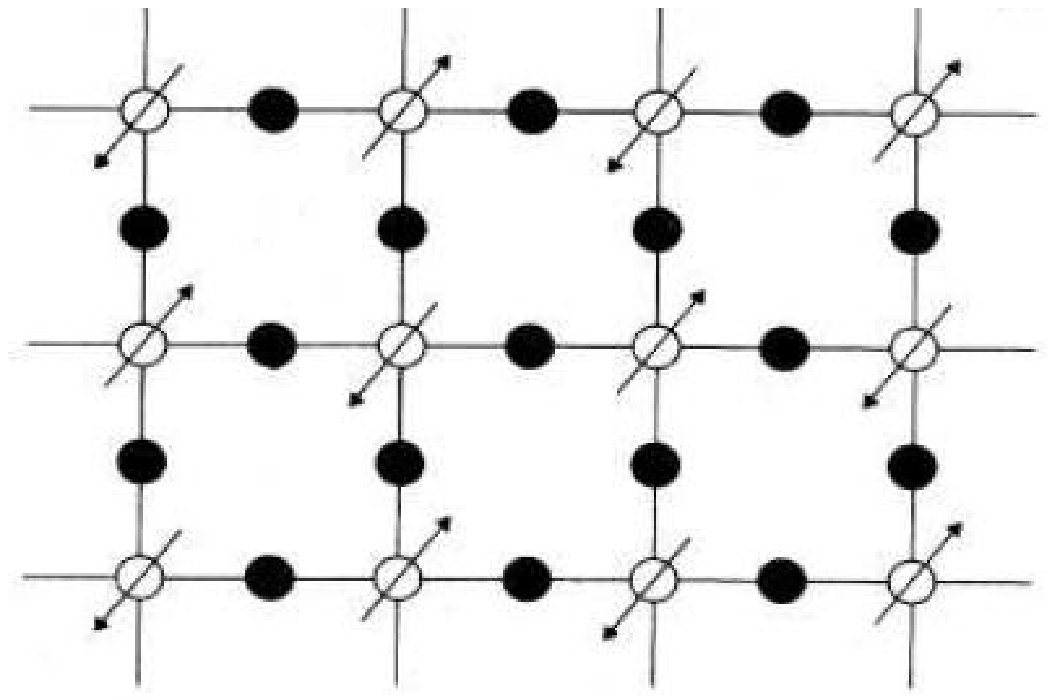}\hspace{10mm}\\
(b)\,\,\,\,\,\,\,\, \includegraphics[scale=0.8]{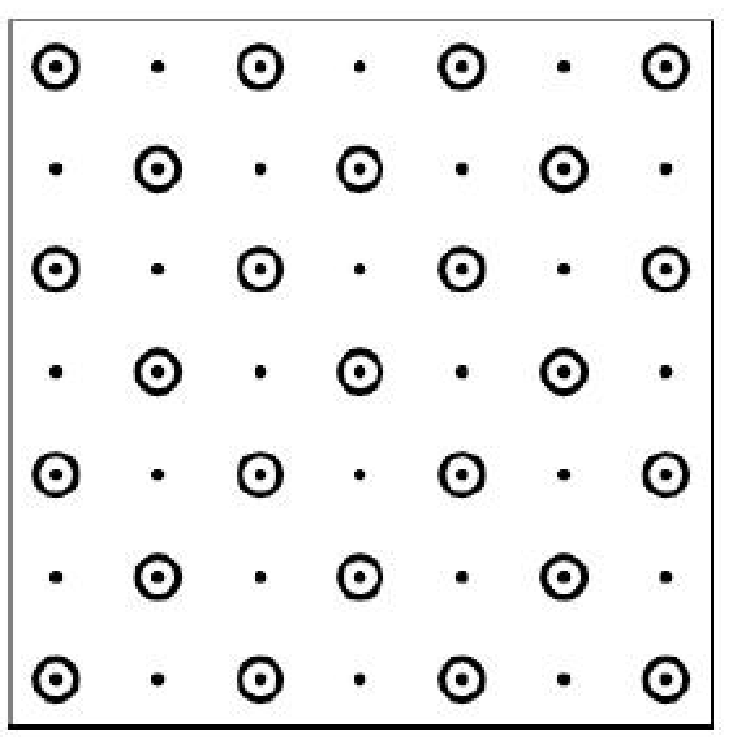}
\par\end{centering}
\vspace{.2cm}
\caption{{\protect\small (a) Estructura magn\'{e}tica del }$La_{2}CuO_{4}%
${\protect\small {} en sus planos }bidimensionales $Cu$-$O$.{\protect\small {}
El }$Cu${\protect\small {} y el }$O${\protect\small {} se representan mediante
c\'{\i}rculos abiertos y cerrados respectivamente. (b) Red puntual asociada al
modelo de los planos }$Cu$-$O${\protect\small . En la b\'{u}squeda de
propiedades de correlaciones fuertes del gas de electrones fue \'{u}til
liberar restricciones de simetr\'{\i}a al separar la red puntual absoluta en
dos subredes representadas con puntos coincidentes con los sitios de cobre.}}%
\label{antif}
\end{figure}
En su fase normal el $La_{2}CuO_{4}$ es antiferromagn\'{e}tico (Fig. \ref{antif} a) y
se tiene que la invarianza de traslaci\'{o}n que lleva de un $Cu$ a un $Cu$
vecino cercano se rompe. Por ese motivo en \cite{key-1,key-2,key-3} se
consider\'{o} que los estados de una part\'{\i}cula en el tratamiento HF
pudieran romper la invarianza de traslaci\'{o}n. Por tanto el estado
f\'{\i}sico que describe los orbitales de HF debe ser invariante  solamente ante las
traslaciones discretas que transforman una subred en ella misma (Fig. \ref{antif} b),
pero no ante las que transformen una subred en la otra. Este grupo de
traslaciones es un subgrupo del conjunto de simetr\'{\i}as del cristal
original y por ende su representaci\'{o}n en el espacio inverso $\mathbf{k}$,
debe ser m\'{a}s reducida en n\'{u}mero de estados.
Teniendo en cuenta lo mencionado antes, se definen
cada una de las dos subredes puntuales $r=1\,\, o\,\,2$ en
la forma:
\begin{equation}
\mathbf{R}^{(r)}=\sqrt{2}n_{1}p\,\mathbf{q}_{1}+\sqrt{2}n_{2}p\,\mathbf{q}%
_{2}+\mathbf{q}^{(r)}, \label{eq:11}%
\end{equation}
\[
\mathbf{q}^{(r)}=\left\{
\begin{array}
[c]{c}%
\mathbf{0},\,\,\,\,\,\,\,\, si\,\, r=1,\\
p\,\mathbf{e}_{x_{1}},\,\, si\,\, r=2,
\end{array}
\right.
\]
donde $\mathbf{q}_{1}$ y $\mathbf{q}_{2}$ son los versores base de ambas subredes.

As\'{\i} pues, las soluciones que buscamos son autofunciones del grupo de
traslaciones discretas $\hat{T}_{\mathbf{R}^{(1)}}$, que transforman una
subred en s\'{\i} misma:
\begin{equation}
\hat{T}_{\mathbf{R}^{(1)}}\phi_{\mathbf{k},l}=\exp(i\mathbf{k}\cdot
\mathbf{R}^{(1)})\phi_{\mathbf{k},l}. \label{eq:12}%
\end{equation}

\begin{figure}
\begin{centering}
\includegraphics[scale=0.6]{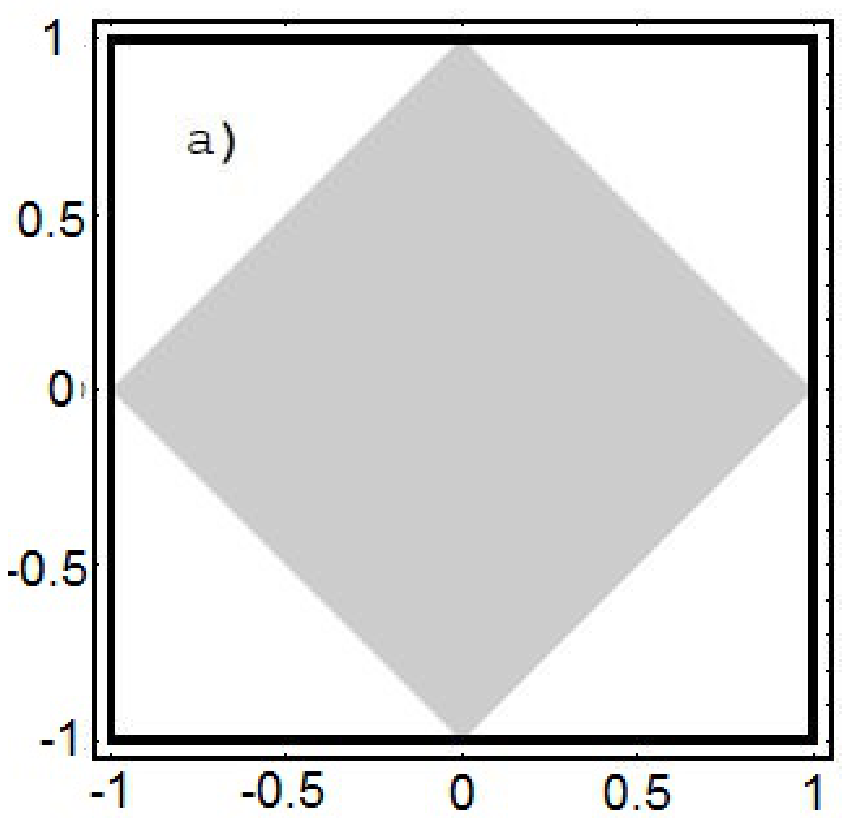}
\hspace{10mm}\includegraphics[scale=0.6]{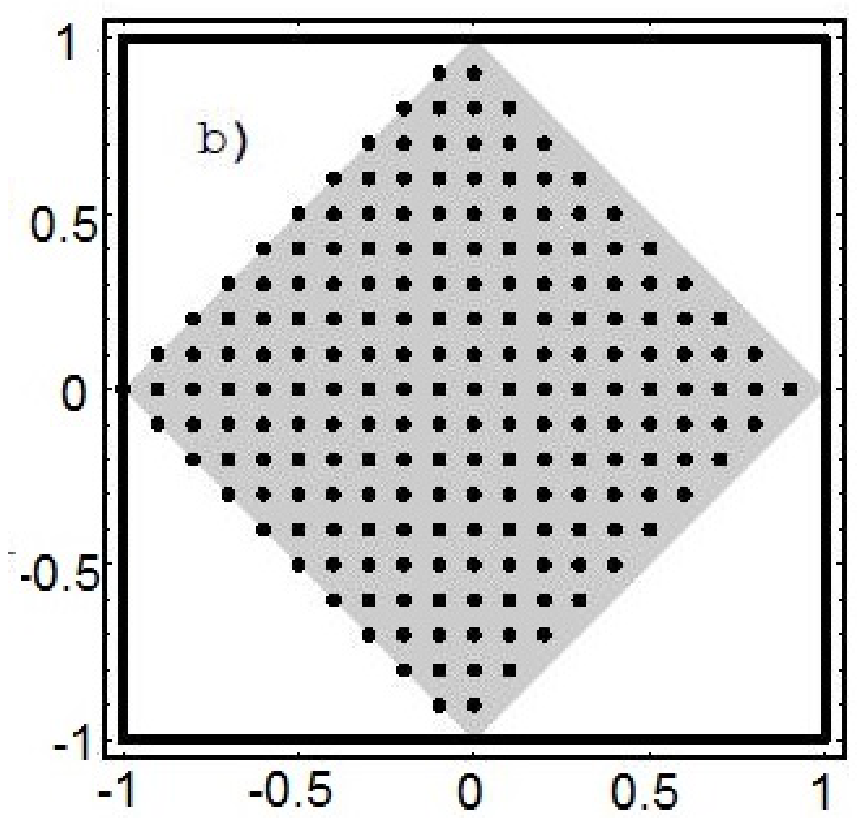}
\par\end{centering}
\caption{{\protect\small (a) Se muestra en gris la zona de Brillouin asociada
al grupo} $\hat{T}_{\mathbf{R}^{(1)}}$ {\protect\small para una red puntual
infinita. (b) La red de puntos muestra el car\'{a}cter discreto de la zona de
Brillouin asociada a }$\hat{T}_{\mathbf{R}^{(1)}}${\protect\small {} cuando la
red puntual es finita con condiciones peri\'{o}dicas en sus fronteras. La
escala unidad significa }$\frac{\pi}{p}${\protect\small {} .}}
\label{brillouin}
\end{figure}

Si la red puntual fuese infinita, la zona de Brillouin asociada a $\hat
{T}_{\mathbf{R}^{(1)}}$ ser\'{\i}a la zona sombreada en la Figura \ref{brillouin}a,
mientras que el cuadrado continente representa la asociada al grupo de
traslaciones en la red total. Dado que en un estudio n\'umerico resulta
imposible considerar  la red
infinita, escogeremos dentro ella una red de estados
$\mathbf{k}$ que implementen las condiciones de  periodicidad de las $\phi_{\mathbf{k},l}$ en las
fronteras de la red total $x_{1}=-Lp$ y $Lp$, $x_{2}=-Lp$ y $Lp$ (Fig. \ref{brillouin}b
). Esta condici\'on  determina los valores
\[
\mathbf{k}=\left\{
\begin{array}
[c]{c}%
\frac{2\pi}{Lp}(n_{x_{1}}\mathbf{e}_{x_{1}}+n_{x_{2}}\mathbf{e}_{x_{2}})\\
con\,\, n_{x_{1}},n_{x_{2}}\in Z \\
y\,\,-\frac{L}{2}\leq n_{x_{1}}\pm n_{x_{2}}<\frac{L}{2}%
\end{array}
\right.  . \label{7}
\]

Luego el n\'{u}mero de elementos en este subgrupo de traslaciones es la mitad de la cantidad de elementos que hay en el grupo de traslaciones en la red puntual absoluta. Trabajemos ahora en una base que
cumpla con (7).
 Definamos as\'i la base tight-binding en aproximaci\'{o}n de una banda%
\begin{equation}
\varphi_{\mathbf{k}}^{(r,\sigma_{z})}(\mathbf{x},s)=\sqrt{\frac{2}{N}%
}u^{\sigma_{z}}(s)\sum_{\mathbf{R}^{(r)}}\exp(i\mathbf{k}\cdot\mathbf{R}%
^{(r)})\varphi_{0}(\mathbf{x}-\mathbf{R}^{(r)}), \label{eq:13}%
\end{equation}
\[
\hat{\sigma}_{z}u^{\sigma_{z}}=\sigma_{z}u^{\sigma_{z}},
\]
\begin{equation}
\varphi_{0}(\mathbf{x})=\frac{1}{\sqrt{\pi a^{2}}}\exp(-\frac{\mathbf{x}^{2}%
}{2a^{2}}),\,\,\, a\ll p, \label{eq:14}%
\end{equation}
donde $N$ es la cantidad de electrones en el gas din\'{a}mico y $\hat{\sigma
}_{z}$ el operador de proyecci\'{o}n de esp\'{\i}n en la direcci\'{o}n $z$,
que para nosotros es la perpendicular a los planos $CuO_{2}$; $\sigma
_{z}=-1\,\, o\,\,1$, sus autovalores; $r=1\,\, o\,\,2$ es el \'indice de cada
una de las subredes. En la aproximaci\'{o}n de solapamiento peque\~{n}o entre
vecinos cercanos, o sea de subredes diferentes, solo se pierde el car\'{a}cter
ortogonal de elementos correspondientes a distintas subredes con la misma
cuantizaci\'{o}n de esp\'{\i}n. Sin embargo, la ortogonalizaci\'{o}n de
elementos distintos correspondientes a la misma subred, as\'{\i} como la norma
unidad de todo elemento, se mantiene, ya que solo implican solapamiento entre
vecinos no cercanos.

Los orbitales de Wannier aqui  propuestos $\varphi_{0}(\mathbf{x}-\mathbf{R}^{(r)})$  representan
la amplitud de probabilidad de encontrar un electr\'{o}n
en el sitio $\mathbf{R}^{(r)}$, o sea, en la base $CuO_{2}$ que el representa.
Consideramos aqu\'i solamente una aproximaci\'{o}n de un banda, ya que nuestra intenci\'{o}n
no fue realizar un estudio exacto del problema.  Pretendemos  solo considerar  sus
rasgos principales en v\'{\i}as de obtener soluciones que reflejen las
propiedades f\'{\i}sicas del material. Siguiendo este
principio hemos  considerado que el potencial efectivo sobre cada electr\'{o}n del
gas, es cuadr\'{a}tico en las vecindades de cada sitio $Cu$, y como referimos
anteriormente, fuertemente confinante a la celda $CuO_{2}$. Esta suposici\'on   justifica
 la forma Gausiana seleccionada antes  para la funci\'on  de Wannier.

\section{Soluci\'on de campo medio }

En esta secci\'on  se presenta el problema
matricial equivalente, que resulta de proyectar las ecuaciones de HF
asociadas al sistema,  en la base Tight-Binding (\ref{eq:13}) definida en el cap\'{\i}tulo anterior.
Se reajustan  aqu\'i los  par\'{a}metros
del modelo al fijar en la forma siguiente: el ancho de banda  de la soluci\'on  paramagn\'{e}tica met\'{a}lica al valor de la banda de
Matheiss $3.8$ $eV$, el gap del
estado b\'{a}sico AFA a $2.0$ $eV$ y el valor de la constante diel\'{e}ctrica
de $21$, que es cercano al medido experimentalmente.

Sean los estados de HF de una sola part\'icula a determinar,  escritos en la forma
\begin{equation}
\phi_{\mathbf{k},l}(\mathbf{x},s)=\sum_{r,\sigma_{z}}B_{r,\sigma_{z}%
}^{\mathbf{k},l}\varphi_{\mathbf{k}}^{(r,\sigma_{z})}(\mathbf{x},s),
\label{eq:15}%
\end{equation}
donde $\varphi_{\mathbf{k}}^{(r,\sigma_{z})}(\mathbf{x},s)$ son los elementos de la base
tight-binding antes definida,
$\mathbf{k}$ el vector de onda asociado al espacio rec\'{\i}proco de la red
absoluta y $l$ es el \'indice   de los restantes n\'{u}meros cu\'{a}nticos
necesarios para precisar el estado de una part\'icula  en cuesti\'{o}n.

Las ecuaciones de HF sin restricciones algunas sobre los orbitales de una part\'icula
 fueron obtenidas  por Dirac \cite{Dirac} y sus expresiones b\'asicas aplicadas al sistema en consideraci\'on
son discutidase en detalle en las referencias \cite{key-1,key-2,key-3}.
Despu\'es de proyectar las ecuaciones de HF escritas en la representaci\'on de coordendas
 en la base de funciones (\ref{eq:13}),  se puede obtener  la  siguiente versi\'on  matricial del problema
autoconsistente (para los detalles ver \cite{key-1,key-2,key-3}):
\begin{equation}
\left[  E_{\mathbf{k}}^{0}+\tilde{\chi}(G_{\mathbf{k}}^{C}-G_{\mathbf{k}}%
^{i}-F_{\mathbf{k}})\right]  \cdot B^{\mathbf{k},l}=\tilde{\varepsilon}%
_{l}(\mathbf{k})I_{\mathbf{k}}\cdot B^{\mathbf{k},l}, \label{eq:16}%
\end{equation}
donde las constantes:
\begin{equation}
\tilde{\chi}=\frac{me^{2}a^{2}}{4\pi\hbar^{2}\epsilon\epsilon_{0}p}, \, \,
\tilde{\varepsilon}_{l}(\mathbf{k})=\frac{ma^{2}}{\hbar^{2}}\varepsilon
_{l}(\mathbf{k}), \label{eq:18}%
\end{equation}
son adimensionales, al igual que todos los par\'{a}metros impl\'{\i}citos en
la definici\'{o}n de las matrices:
\[
E_{\mathbf{k}}^{0}=\left\Vert E_{\mathbf{k},(t,r,\alpha_{z},\sigma_{z})}%
^{0}\right\Vert _{4\times4}\,,\,\, G_{\mathbf{k}}^{C}=\left\Vert
G_{\mathbf{k},(t,r,\alpha_{z},\sigma_{z})}^{C}\right\Vert _{4\times4}\,,
\]
\[
G_{\mathbf{k}}^{i}=\left\Vert G_{\mathbf{k},(t,r,\alpha_{z},\sigma_{z})}%
^{i}\right\Vert _{4\times4}\,,\,\, F_{\mathbf{k}}=\left\Vert F_{\mathbf{k}%
,(t,r,\alpha_{z},\sigma_{z})}\right\Vert _{4\times4}\,,
\]
\[
I_{\mathbf{k}}=\left\Vert I_{\mathbf{k},(t,r,\alpha_{z},\sigma_{z}%
)}\right\Vert _{4\times4}\,.
\]
Los diferentes t\'erminos que participan en esta ecuaci\'on son: el potencial peri\'{o}dico del medio $W_{\gamma}$,
los t\'{e}rminos de Coulomb $G_{\mathbf{k}}^{C}$ y de intercambio $G_{\mathbf{k}}^{i}$, el potencial de interacci\'{o}n
con el fondo neutralizante $F_{b}$ y la matriz de solapamiento $I_{\mathbf{k}}$ entre vecinos cercanos,
respectivamente. La forma de los elementos matriciales se da expl\'icitamente  en los
Ap\'{e}ndices de \cite{key-1,key-3}. En esta  representaci\'{o}n la energ\'{\i}a HF a $T=0\, K$ del sistema y la
condici\'{o}n de normalizaci\'{o}n de la funci\'{o}n de onda de cada estado
uniparticular,  toman la forma:
\begin{eqnarray}%
E^{HF} & = & \sum_{\mathbf{k},l}\Theta_{(\tilde{\varepsilon}_{F}%
-\tilde{\varepsilon}_{l}(\mathbf{k}))}.[\tilde{\varepsilon}_{l}(\mathbf{k}%
)-\frac{\tilde{\chi}}{2}B^{\mathbf{k},l*}.(G_{\mathbf{k}}^{C}-G_{\mathbf{k}%
}^{i}).B^{\mathbf{k},l}], \nonumber
\\
1 & = & B^{\mathbf{k},l*}.I_{\boldsymbol{k}}.B^{\mathbf{k},l}.
\label{eq:19}
\end{eqnarray}

El sistema (\ref{eq:16}) es no lineal en las variables $B_{r,\sigma_{z}}%
^{\mathbf{k},l}$, que son las cuatro componentes de cada vector $B^{\mathbf{k}%
,l}$, y que se interpretan como  amplitudes de probabilidad de
encontrar al electr\'{o}n en el estado $(\mathbf{k},l)$, en la subred $r$, con
cuantizaci\'{o}n $\sigma_{z}$ del esp\'{\i}n a lo largo del eje $z$. Con
vistas a resolverlo num\'{e}ricamente por el m\'{e}todo de iteraciones
sucesivas, es conveniente premultiplicarlo por $I_{\boldsymbol{k}}$ para cada
$\mathbf{k}$. N\'{o}tese que para cada $\mathbf{k}$ se obtendr\'{a}n cuatro
autovalores ($l=1,2,3,4$), o lo que es igual, cuatro bandas en la ZB. La invariancia de traslaci\'on sobre las
 subredes implica que en la representaci\'{o}n (\ref{eq:13}) el potencial de
interacci\'{o}n HF y en general el operador de Fock del problema de HF, son
diagonales en bloque respecto a los estados $\mathbf{k}$. Esto  es
consecuencia directa de su conmutaci\'{o}n con el grupo de traslaciones discreta
reducido.

\subsection{Ajuste de los par\'{a}metros libres}

En la presente subsecci\'on ajustamos los par\'{a}metros libres introducidos en el
modelo, $\epsilon$: constante diel\'{e}ctrica del medio efectivo; $m:$ masa
efectiva del medio; $\tilde{a}$: radio en que superviven los orbitales de
Wannier gaussianos; $\tilde{\gamma}$: amplitud de probabilidad de salto de un
sitio a otro cercano fijada por el medio efectivo y $\tilde{b}$: radio en que
supervive la densidad de carga asociada al medio neutralizante. Con este fin y
apoyados en los resultados obtenidos en las referencias
\cite{key-1,key-2,key-3}, buscamos fijar simult\'{a}neamente:  el ancho de banda
de Matheiss a $3.8$$\,\, eV$ (tal y como muestra el perfil de dispersi\'{o}n
de la Figura \ref{matheiss}),  el gap del estado normal antiferromagn\'{e}tico aislante del
$La_{2}CuO_{4}$ a $2.0\,\,$$eV$ y el valor observado de su constante
diel\'{e}ctrica igual a $21$.

Con vistas a obtener un estado  paramagn\'{e}tico met\'{a}lico que pudiera describir
la banda calculada por  Mathiess,  se
busc\'{o} primeramente la soluci\'on HF del porblema en un espacio de funciones de Bloch del grupo maximal de
traslaciones, esto es, que dejan invariante la red puntual absoluta. La base
de Bloch Tight-Binding de una banda para este problema tiene la forma adoptada
en \cite{key-1,key-2,key-3} ,
\begin{equation}
\varphi_{\mathbf{Q}}^{\sigma_{z}}(\mathbf{x},s)=\sqrt{\frac{1}{N}}%
u^{\sigma_{z}}(s)\sum_{\mathbf{R}}\exp(i\mathbf{Q}\cdot\mathbf{R})\varphi
_{0}(\mathbf{x}-\mathbf{R}), \label{eq:20}%
\end{equation}
donde los momenta $\mathbf{Q}$ que aparecen se definen por
\[ \mathbf{Q}=\left\{
\begin{array}
[c]{c}%
\frac{2\pi}{Lp}(n_{x_{1}}\mathbf{e}_{x_{1}}+n_{x_{2}}\mathbf{e}_{x_{2}})\\
con\,\, n_{x_{1}},n_{x_{2}}\in Z\\
-\frac{L}{2}\leq n_{x_{1}},n_{x_{2}}<\frac{L}{2}%
\end{array}
\right\}.
\]

\begin{figure}
\begin{centering}
(a)\includegraphics[scale=0.45]{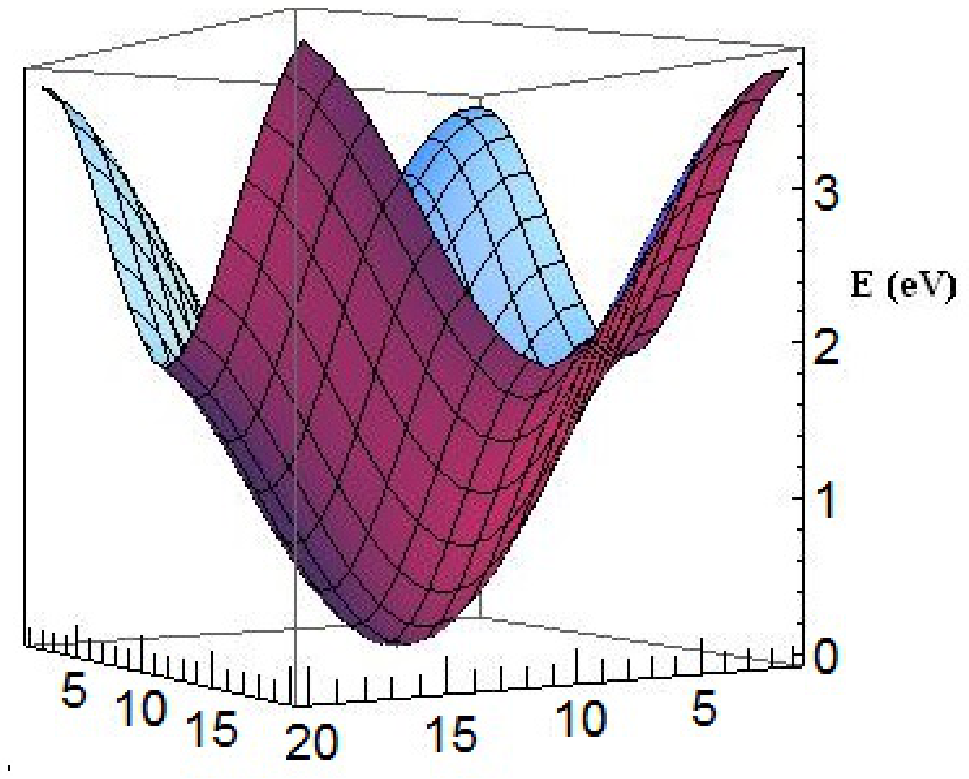}\hspace{10mm}
(b)\includegraphics[scale=0.45]{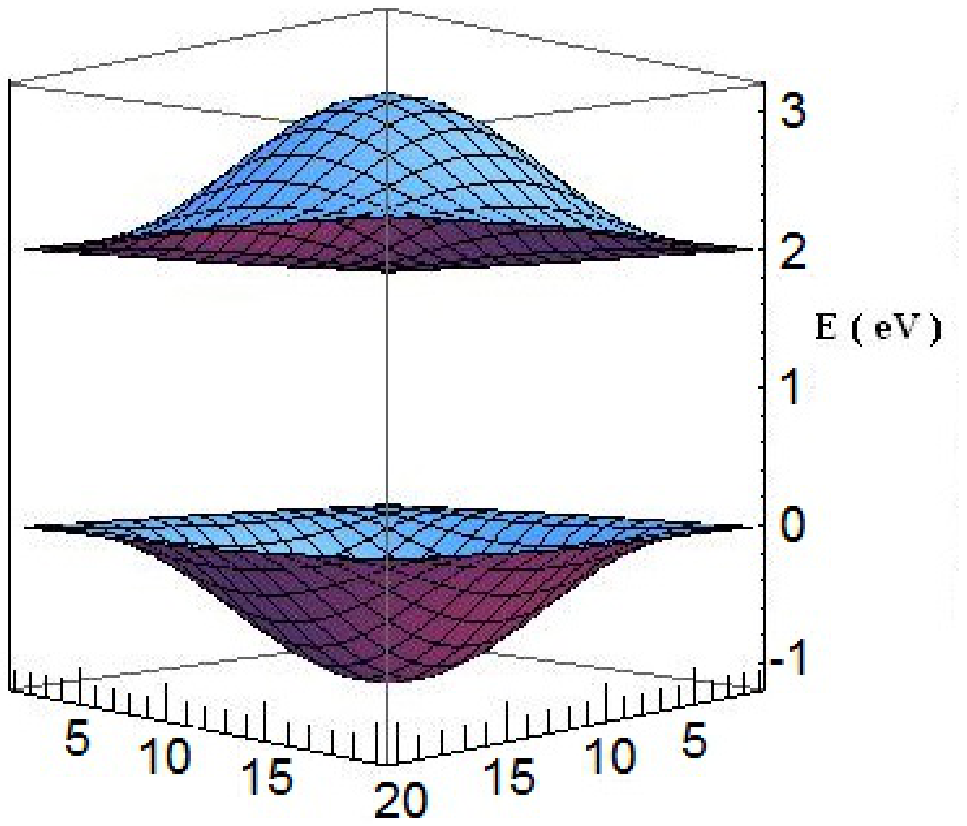}
\par\end{centering}
\caption{{\protect\small (a) Banda de energ\'{\i}a doblemente degenerada
paramagn\'{e}tica y met\'{a}lica. (b) Banda de energ\'{\i}a AFA. The unit of the integer number scales shown in the plot is defined as the length of the side of the Brillouin cell associated to the sublattices: $\sqrt{2}\frac {\pi}{p}$,  after divided by 20. The same convention  is employed for other plots shown below.}}%
\label{twofig}
\end{figure}
Adem\'{a}s $N=L\times L$ y $\mathbf{R}$ son, respectivamente, la cantidad de
celdas base en la red puntual absoluta y las coordenadas de la misma.
Sean los estados de Bloch que buscamos en la base antes mencionada expresados
en la forma:
\begin{equation}
\phi_{\mathbf{Q},l}(\mathbf{x},s)=\sum_{\sigma_{z}}B_{\sigma_{z}}%
^{\mathbf{Q},l}\varphi_{\mathbf{Q}}^{\sigma_{z}}(\mathbf{x},s). \label{eq:21}%
\end{equation}
El problema matricial equivalente para las $B^{\mathbf{Q},l}$ es esta vez de
segundo orden para cada estado $(\mathbf{Q},l)$, es decir, ser\'{a}n vectores
de 2 componentes. As\'{\i} pues, en forma an\'{a}loga a como se deriv\'o (\ref{eq:16}), se puede obtener el
sistema de ecuaciones de HF en la forma:
\begin{equation}
\left[  E_{\mathbf{Q}}^{0}+\tilde{\chi}(G_{\mathbf{Q}}^{C}-G_{\mathbf{Q}}%
^{i}-F_{\boldsymbol{Q}})\right]  \cdot B^{\mathbf{Q},l}=\tilde{\varepsilon
}_{l}(\mathbf{Q})I_{\mathbf{Q}}\cdot B^{\mathbf{Q},l}, \label{eq:22}%
\end{equation}
que constituyen un conjunto de ecuaciones matriciales no lineales a resolver.

Para comenzar la iteraci\'{o}n se utiliz\'{o} un estado inicial
paramagn\'{e}tico. En la Figura \ref{twofig}a, se muestra la banda paramagn\'{e}tica,
met\'{a}lica y doblemente degenerada obtenida iterativamente en condici\'{o}n
de semillenado, es decir con $N=20\times20$ electrones.

Para el caso aislante antiferromagn\'{e}tico se llev\'{o} a cabo la
soluci\'{o}n del sistema de ecuaciones (\ref{eq:16}) por el m\'{e}todo de iteraciones
sucesivas partiendo siempre de un estado con car\'{a}cter
antiferromagn\'{e}tico. En la Figura \ref{twofig}b, se muestran el perfil adoptado en
condici\'{o}n de semillenado obtenido para la red puntual de $20\times20$
puntos. Evidentemente corresponden a bandas de estados aislantes.

Los estados correspondientes, resultaron ser los m\'{a}s estables, o sea, los
de m\'{a}s baja energ\'{\i}a (HF) entre todos los encontrados.
As\'{\i} los valores de los par\'{a}metros del modelo fijados resultaron ser:
$\epsilon=21$, $m=2.5\,\,$$m_{e}$, $\tilde{a}=0.09$, $\tilde{b}=17.125\cdot
10^{-3}$ y $\tilde{\gamma}=-17.125\cdot10^{-3}$.
Obs\'{e}rvese las coincidencias topol\'{o}gicas entre la banda obtenida y la
banda de conducci\'{o}n presentada en la Figura \ref{matheiss}, en ambas el nivel de
Fermi en la direcci\'{o}n $\Gamma$-$X$ pasa a la mitad de los bordes de banda
correspondiente, mientras en la direcci\'{o}n que descansa a 45 grados
respecto a $\Gamma$-$X$ la roza en su extremo superior. As\'i quedan
argumentadas las bases del modelo de una banda que presentamos y la
elecci\'{o}n de par\'{a}metros realizada.

\section{Transici\'{o}n de fase cu\'{a}ntica}

\begin{figure}
\includegraphics[scale=.4]{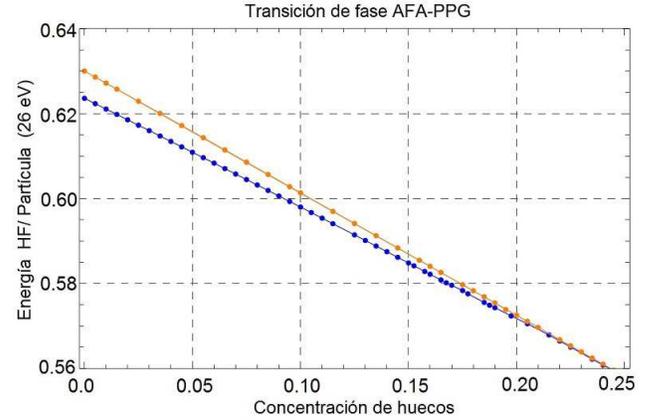}
\caption{ Dependencia de las energ\'{\i}as HF por
part\'{\i}cula en los estados AFA y PPG con respecto al dopaje con huecos.}%
\label{transi}
\end{figure}

\begin{figure}
\begin{centering}
a) \includegraphics[scale=0.3]{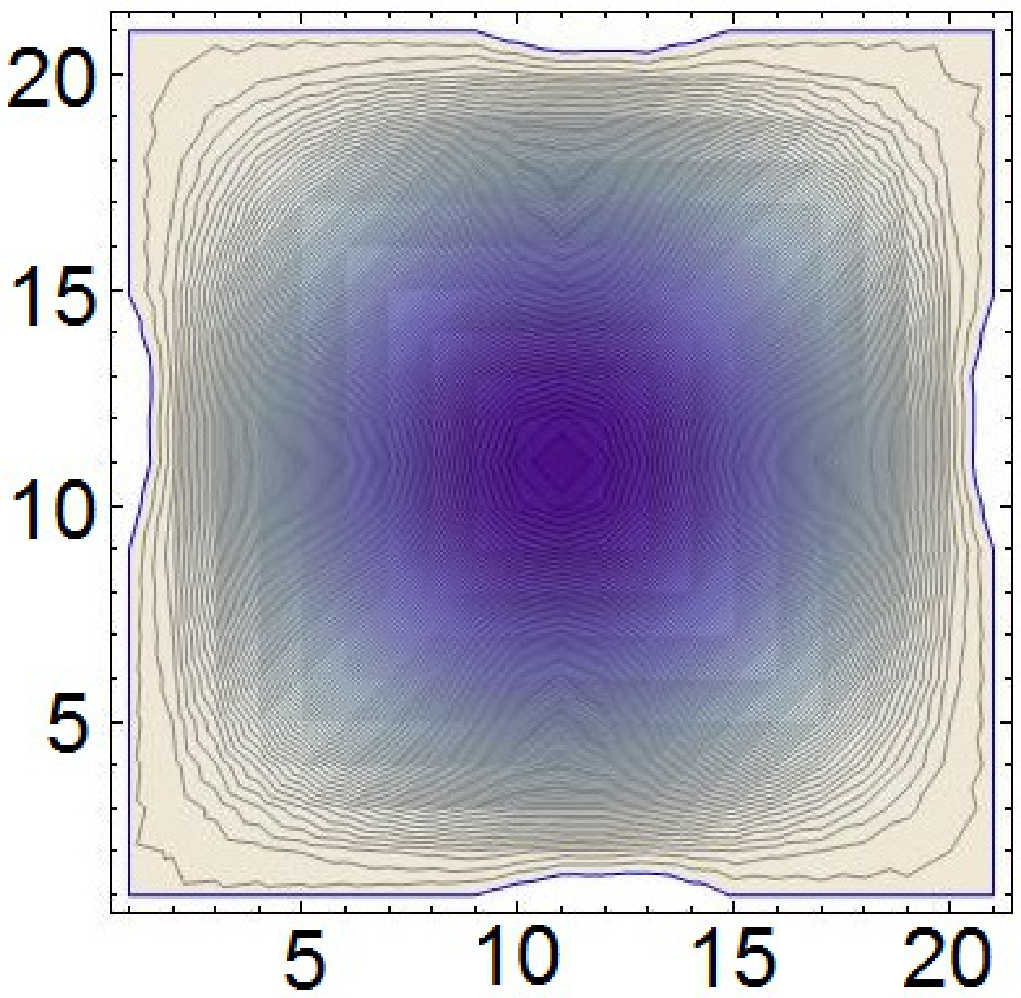}\hspace{10mm}
b) \includegraphics[scale=0.3]{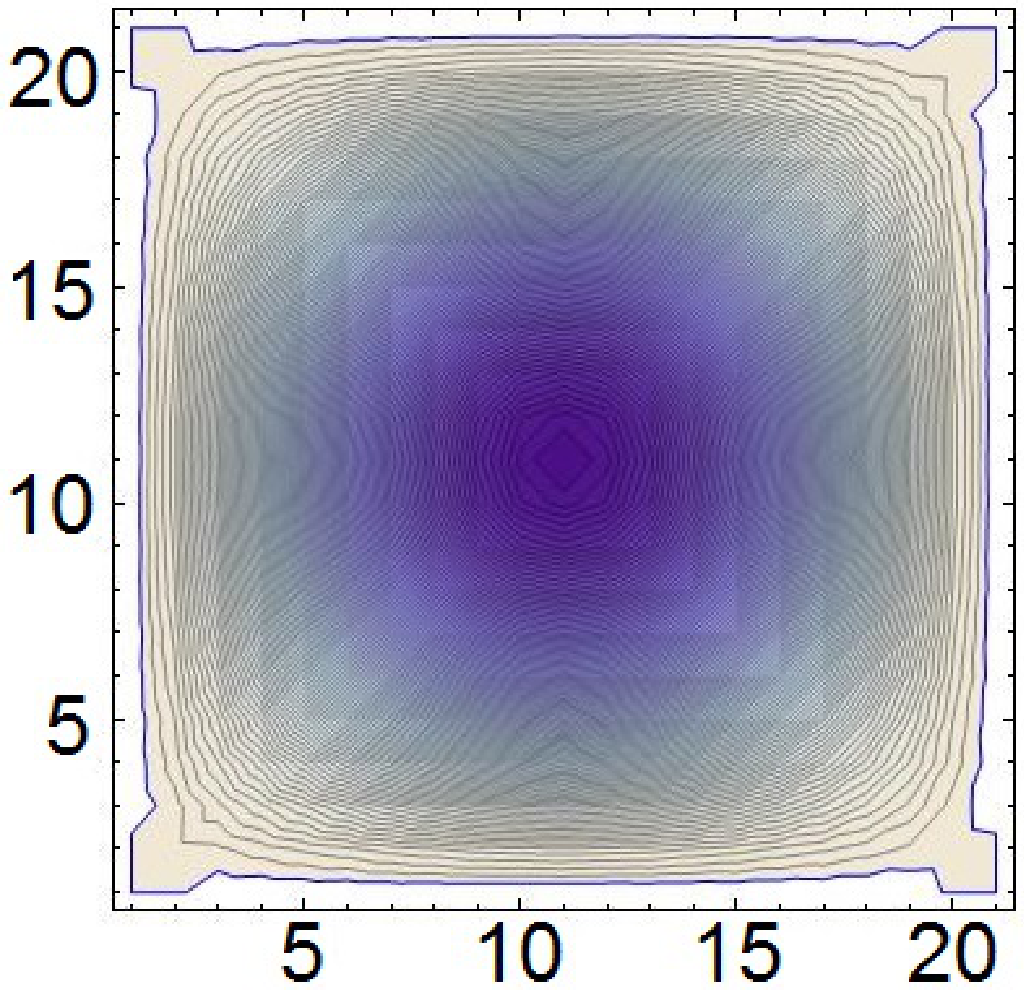}\\
c) \includegraphics[scale=0.3]{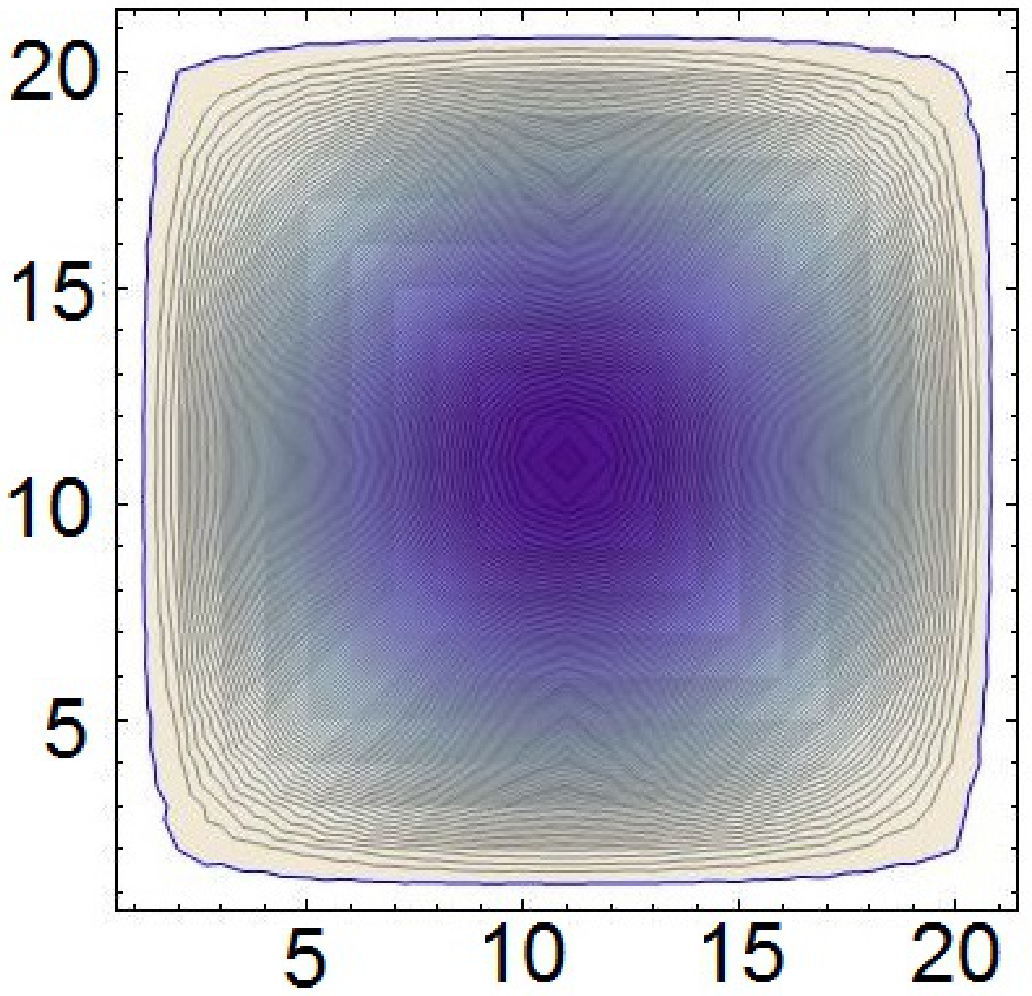}\hspace{10mm}
d) \includegraphics[scale=0.3]{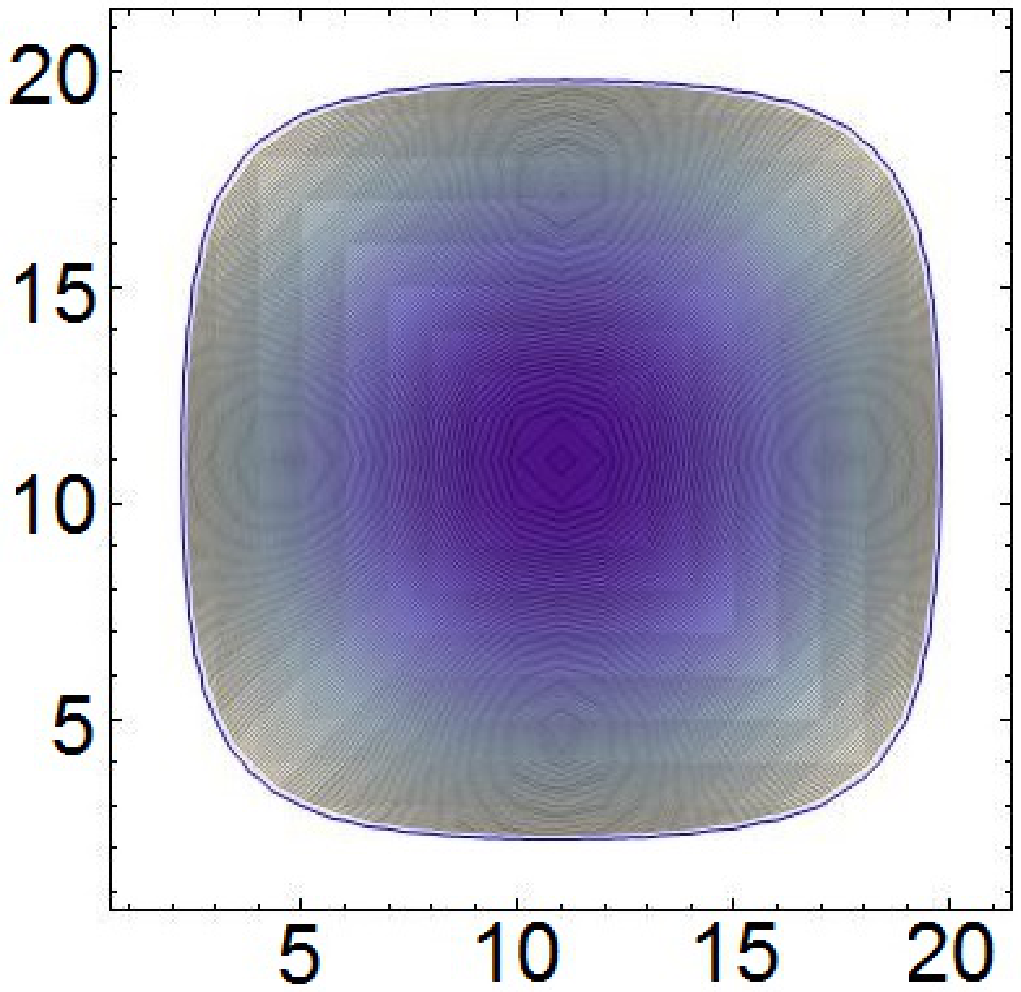}
\par\end{centering}
\caption{{\protect\small Evoluci\'{o}n de la superficie de Fermi a medida que
la concentraci\'{o}n de huecos aumenta a partir del semillenado para el  estado b\'{a}sico AFA. Las
superficies de Fermi mostradas corresponden a los valores de dopaje: }a)
$x=0.02$, b) $x=0.075$, c) $x=0.095$, d) $x=0.3$. Las energ\'ias de los estados uniparticulares
decrecen con el grado de oscuridad en el gr\'afico.{\protect\small {} }}
\label{AF}%
\end{figure}

\begin{figure}
\begin{centering}
a) \includegraphics[scale=0.3]{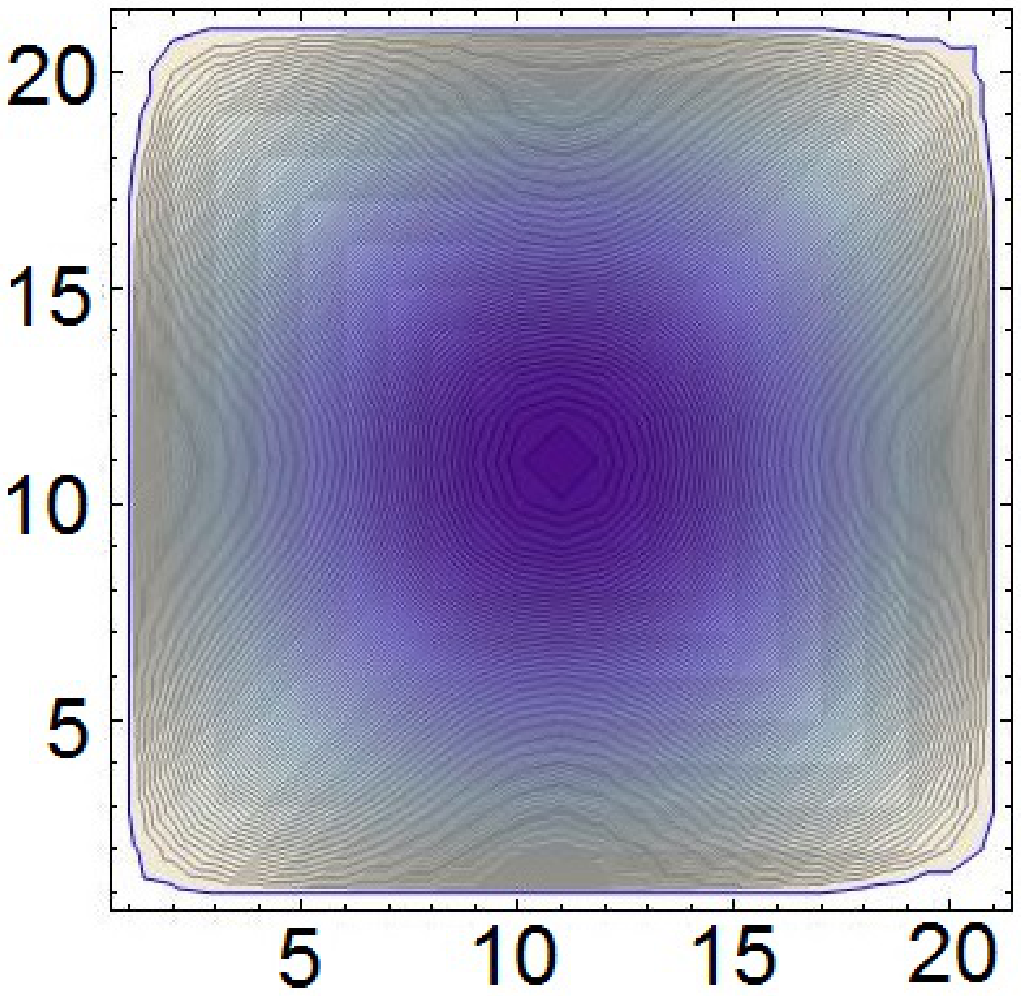}\hspace{10mm}
b) \includegraphics[scale=0.3]{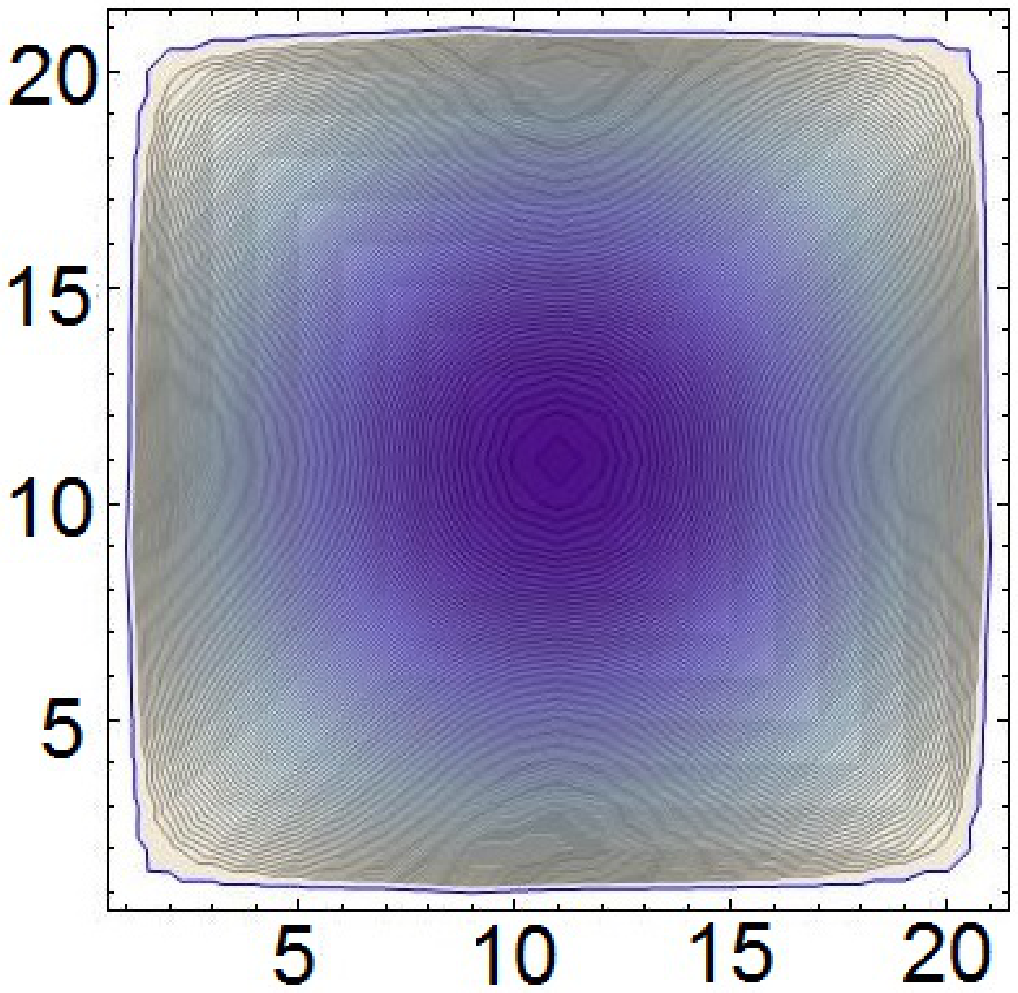}
c) \includegraphics[scale=0.3]{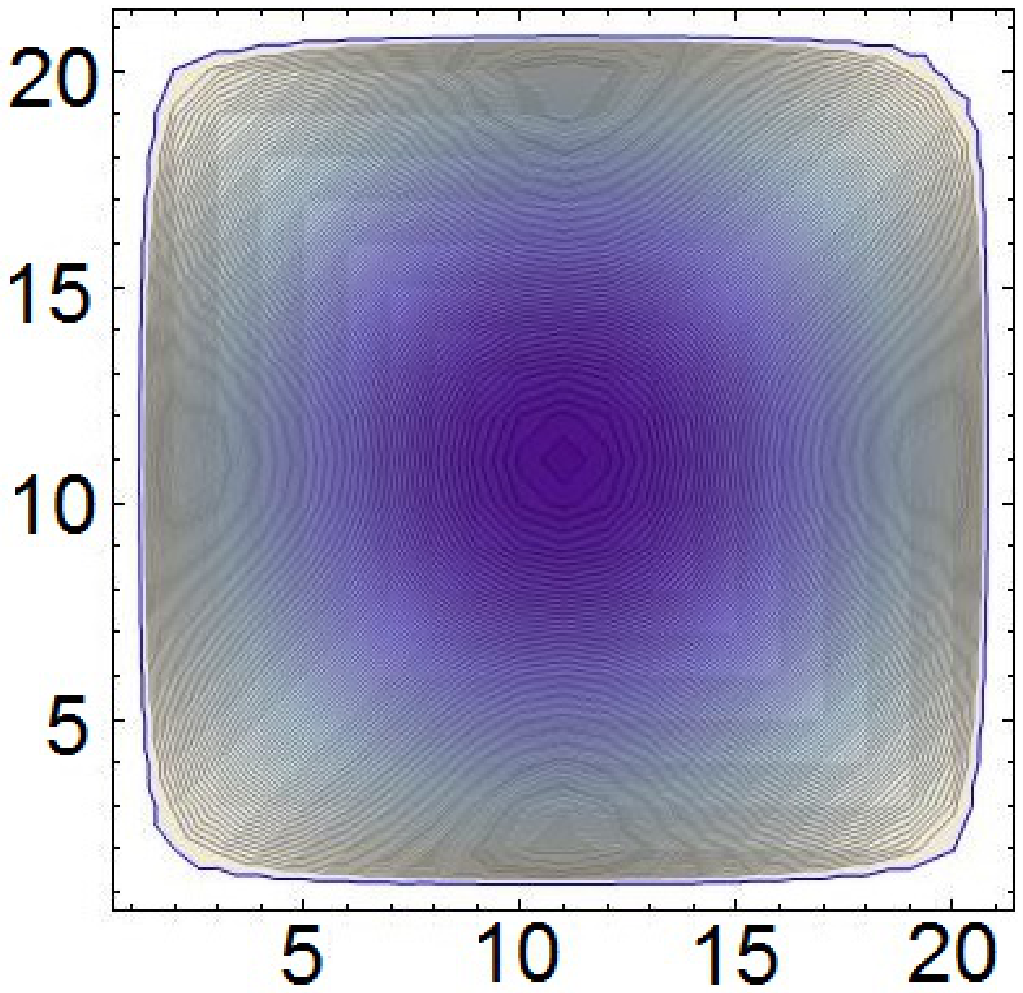}\hspace{10mm}
d) \includegraphics[scale=0.3]{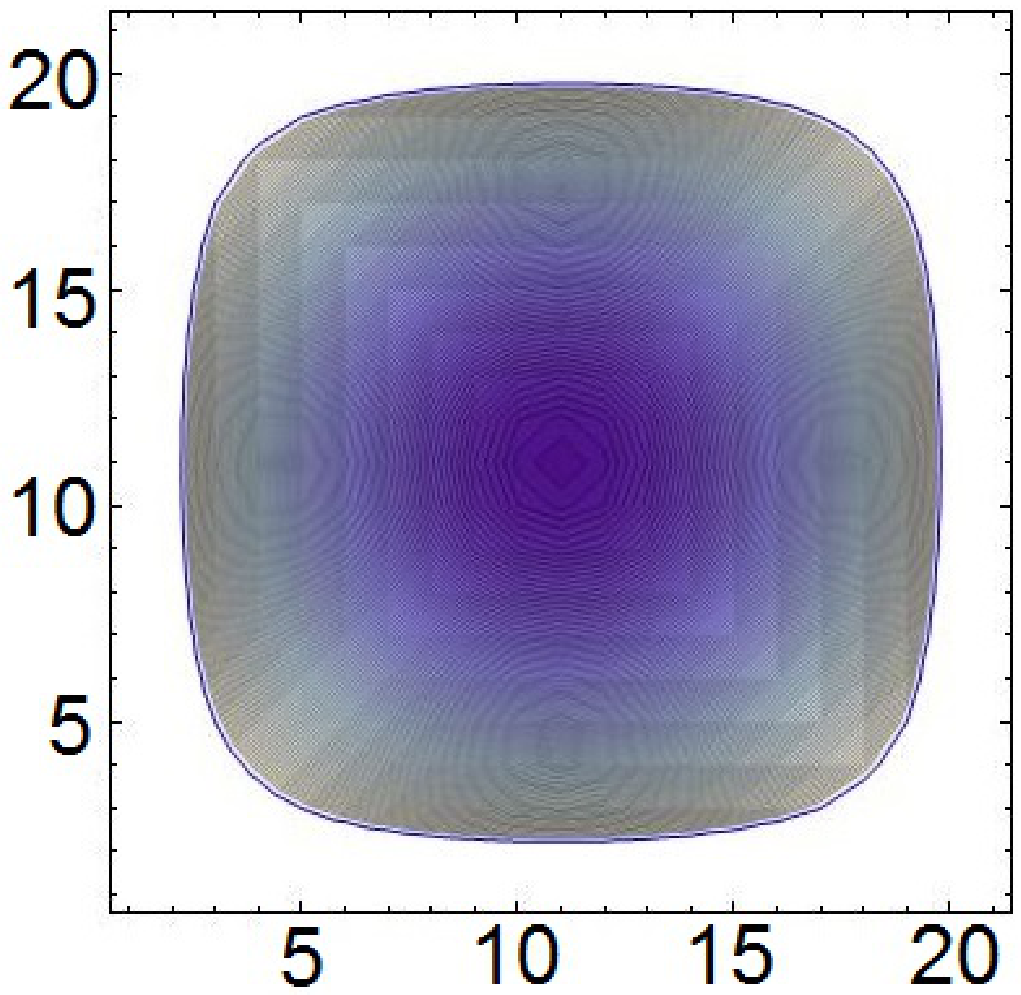}
\par\end{centering}
\caption{{\protect\small Evoluci\'{o}n de la superficie de Fermi a medida que
la concentraci\'{o}n de huecos aumenta a partir del semillenado para el estado excitado PPG. Las
superficies de Fermi mostradas corresponden a los valores de dopaje: }a)
$x=0.02$, b) $x=0.075$, c) $x=0.095$, d) $x=0.3$. Las energ\'ias de los estados uniparticulares
decrecen con el grado de oscuridad en el gr\'afico. }%
\label{PPG}
\end{figure}

La evidencia experimental acerca de la existencia de un punto cr\'{\i}tico
cu\'{a}ntico en el  $La_2CuO_4$ proviene de un
estudio exhaustivo hecho por Jeffery Tallon y John Loram acerca de las
propiedades f\'{\i}sicas de la fase de pseudogap a partir de un gran cuerpo de
experimentos de termodin\'{a}mica que ellos mismos realizaron \cite{key-38}.
Encontraron que el pseudogap se caracteriza por una energ\'{\i}a caracter\'istica que
cae abruptamente a cero en el  dopaje cr\'{\i}tico de 0.19 huecos,  por \'{a}tomo de
cobre en el plano de conducci\'{o}n de un variado n\'{u}mero de cupratos.
Propiedades como la capacidad calor\'{\i}fica electr\'{o}nica cambian
abruptamente en el valor del dopaje cr\'{\i}tico, lo cual indica que puede
existir una transici\'{o}n a temperatura cero entre dos fases distintas.
Pensando en ello, y una vez precisados los estados AFA y PPG en condici\'{o}n
de semillenado, para los valores mejorados de los par\'ametros del modelo,
nos propusimos reconsiderar el estudio presentado en \cite{key-6,key-6b},
para la evoluci\'{o}n con el dopaje de dichos estados a
temperatura cero.
Se calcul\'{o} la energ\'{\i}a HF por part\'{\i}cula al ir variando la
concentraci\'{o}n de huecos en el rango $0\leq x\leq0.25$ para los estados AFA
y PPG mediante la expresi\'{o}n:
\begin{eqnarray}
E^{HF}&=&\sum_{\mathbf{k},l}\Theta_{(\tilde{\varepsilon}_{F}-\tilde{\varepsilon
}_{l}(\mathbf{k}))}[\tilde{\varepsilon}_{l}(\mathbf{k})- \nonumber\\
& & \frac{\tilde{\chi}}%
{2}B^{\mathbf{k},l*}.(G_{\mathbf{k}}^{C}-G_{\mathbf{k}}^{i}).B^{\mathbf{k}%
,l}].
\end{eqnarray}
En la Figura \ref{transi} se muestra como el estado AFA que es el de menor energ\'{\i}a
HF a dopaje cero, evoluciona y se hace degenerado con el PPG para un dopaje
cr\'{\i}tico alrededor de $x_{c}=0.2$. A partir del mencionado valor de
dopaje, los estados AFA y PPG tienden a volverse degenerados, compartiendo as\'{\i}
sus propiedades.
Estudiamos tambi\'en la superficie de Fermi y su dependencia con el dopaje en el
$La_{2-x}Sr_{x}CuO_{4}$ para un amplio rango de concentraci\'{o}n de huecos de
$0\leq x\leq0.3$. En las Figuras \ref{AF} y \ref{PPG}, se muestra la evoluci\'{o}n de la
superficies de Fermi en las fases AFA y PPG en la medida que la
concentraci\'{o}n de huecos aumenta.
Se puede apreciar en el caso AFA, c\'omo
para una peque\~{n}a concentraci\'{o}n de huecos la superficie de Fermi
est\'{a} compuesta de los llamados ``arcos de Fermi'' en el centro de las
caras de la frontera de la zona de Brillouin (ZB).

La longitud de estos arcos aumenta proporcionalmente al dopaje hasta formar unos tipos de bolsas de
huecos (hole pockets) en las esquinas de la ZB. A partir de este punto las
formas de las superficies de Fermi de ambos estados, AFA y PPG, tienden a acercarse.
 De acuerdo con nuestros resultados la superficie de Fermi para $x=0.3$ parece ser
casi cuadrada teniendo una larga porci\'{o}n de rectas paralelas a las caras
de la frontera de la ZB.

Puede  entonces argumentarse que en la medida que se dopa con huecos el
$La_{2}CuO_{4}$, partiendo de su estado normal AFA, su superficie de Fermi
experimenta un cambio dr\'{a}stico al pasar de una superficie de Fermi de
huecos centrada en el nodo ($\pi,\pi$) del espacio rec\'{\i}proco para
$0<x<0.1$ a una superficie de Fermi de electrones centrada en ($0,0$) para
$0.1<x<0.3$.
Debe comentarse que este resultado no coincide exactamente con lo
reportado en la referencia {[}6{]}.
En ese trabajo los huecos pasaron de
estar centrados en las caras a las esquinas de la zona de Brillouin, cerca del
llenado $x_{c}=0.2$. Estimamos que esto pudiera deberse a
la diferencia entre los par\'{a}metros utilizados en el modelo. Un
par\'{a}metro que puede a\'{u}n ajustarse es el ancho de las densidades de
carga de jellium, el cual se asumi\'{o} muy peque\~{n}o. El representar la densidad del  jellium  en
el l\'{\i}mite contrario, es decir, como una
distribuci\'{o}n de carga casi homog\'{e}nea, pudiera cambiar el valor de dopaje en el que la
transici\'{o}n de los huecos de las caras a las esquinas ocurra. Esperamos
estudiar este efecto en las pr\'{o}ximas extensiones del trabajo.

\begin{figure}
\begin{centering}
a) \includegraphics[scale=0.3]{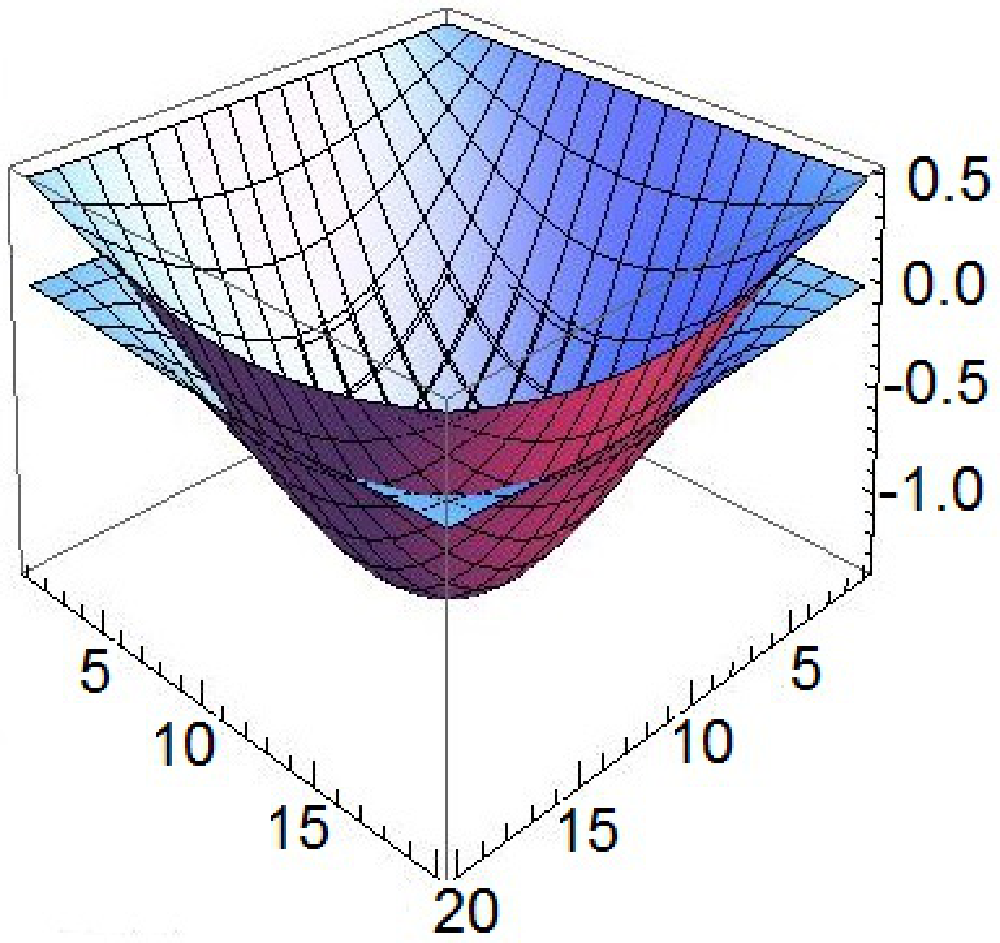}\hspace{10mm}
b) \includegraphics[scale=0.3]{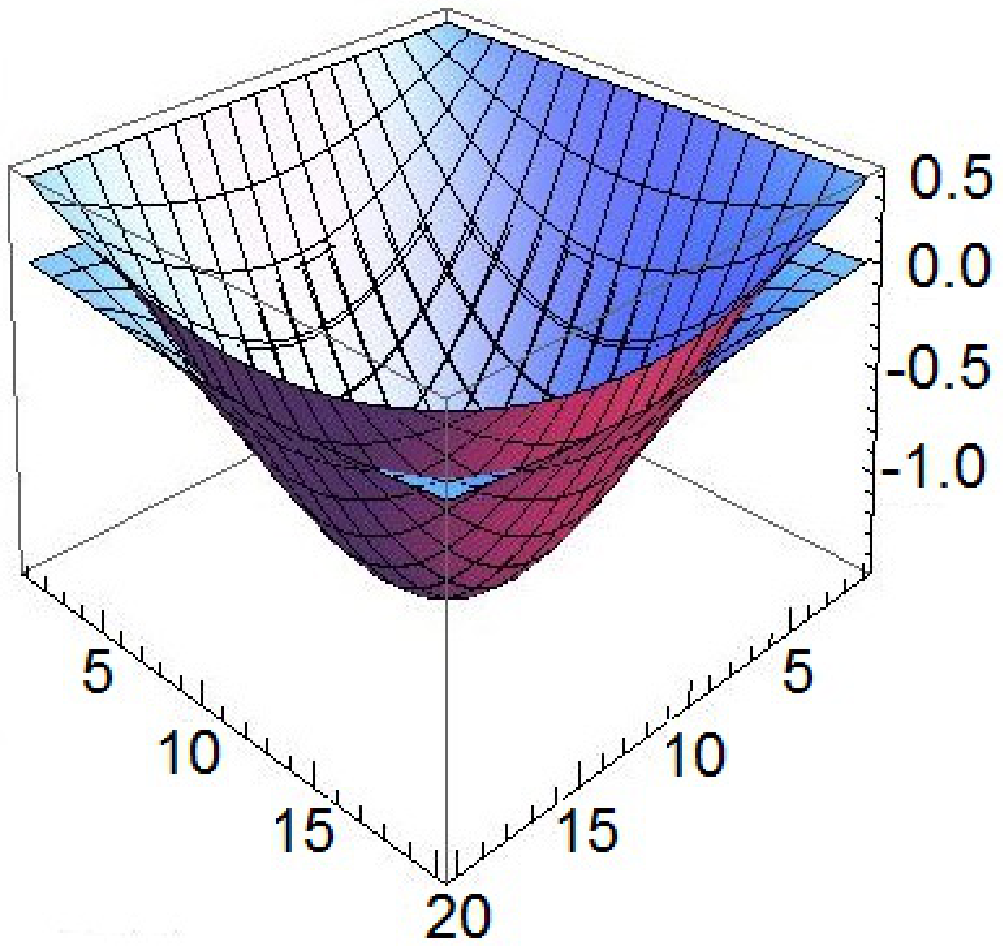}\\
c) \includegraphics[scale=0.3]{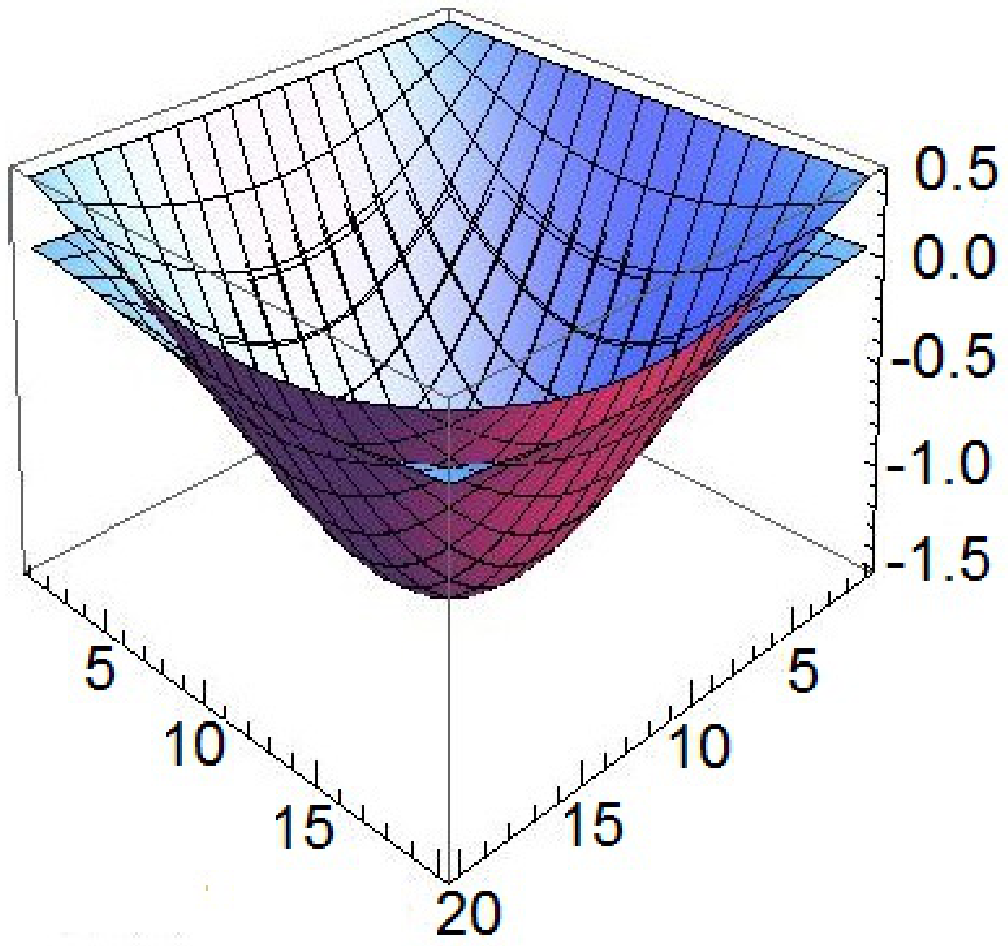}\hspace{10mm}
d) \includegraphics[scale=0.3]{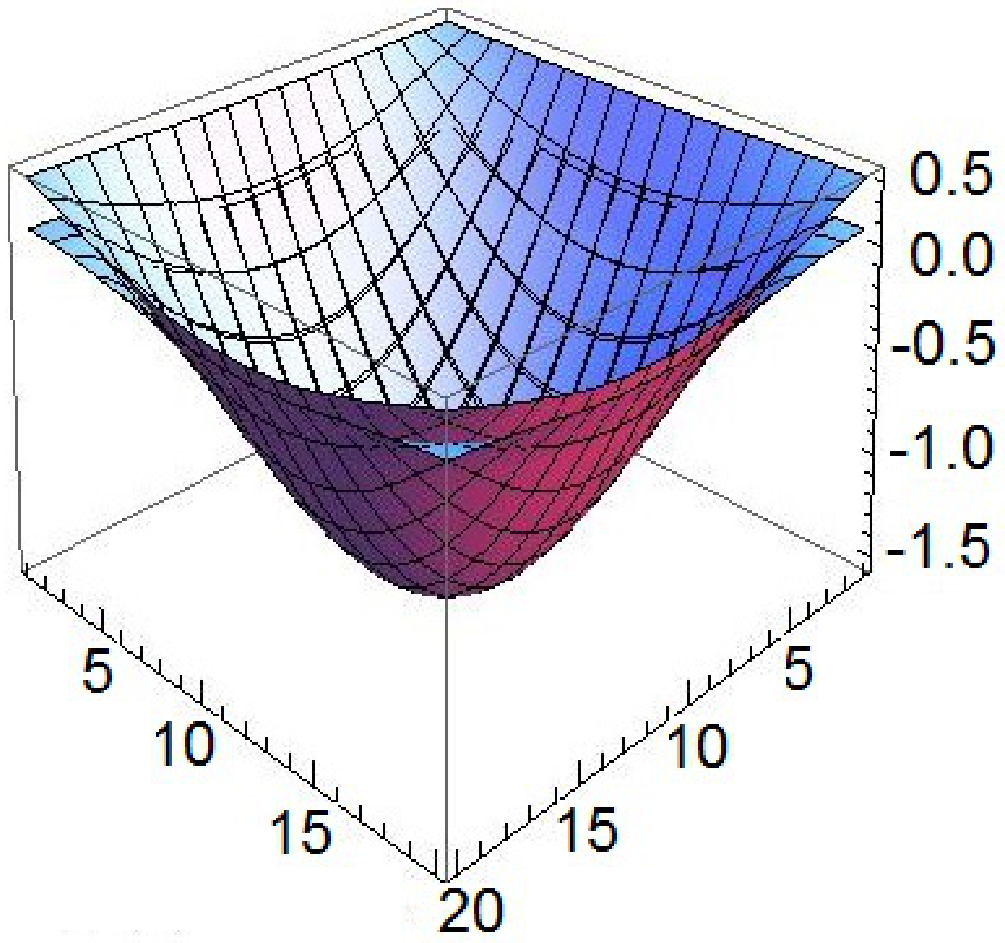}
\par\end{centering}
\caption{{\protect\small Evoluci\'{o}n de las energ\'{\i}as HF de los estados
uniparticulares en las fases AFA y PPG en la medida que aumenta la
concentraci\'{o}n de huecos desde }$x=0${\protect\small {} hasta }%
$x=0.2${\protect\small . El eje vertical refiere a la energ\'{\i}a dada en eV.
Note que se hace coincidir el nivel de Fermi del estado AFA con 0 eV. Las
autoenerg\'{\i}as mostradas corresponden a los valores de dopaje : a) }$x=0$,
b) $x=0.1$, c) $x=0.15$, d) $x=0.2$.}%
\label{deg}
\end{figure}

Tambi\'en representamos  en un marco com\'{u}n las energ\'{\i}as de los estados HF
uniparticulares en las fases AFA y PPG  con vistas a  visualizar  su correspondiente evoluci\'{o}n
en la medida que se dopa el material, y as\'i  tener una idea de la
degeneraci\'{o}n de estados en el sistema. En la Figura \ref{deg} puede observarse
como en condici\'{o}n de semillenado los estados de energ\'{\i}as m\'as bajos
coinciden y la diferencia esencial de energ\'{\i}a se encuentra entre los
estados ubicados en las proximidades de la frontera de la ZB. Al dopar el
material, los huecos crean estados vac\'{\i}os en dicha frontera, tal y c\'omo
mostramos en la evoluci\'{o}n de las superficies de Fermi para ambas fases,
perdiendo as\'{\i} la fase AFA los estados que cargan con un
antiferromagnetismo m\'{a}s fuerte \cite{key-1,key-2,key-6}. Simult\'{a}neamente el
n\'{u}mero de estados con energ\'{\i}as uniparticulares similares va
aumentando al aumentar el dopaje. Las autoenerg\'ias de los orbitales ocupados asociados a los estados AFA y PPG
tienden a coincidir  alrededor de una concentraci\'{o}n de huecos cercana a $x_{c}=0.2$, donde se
ha perdido el orden antiferromagn\'{e}tico \cite{key-6}.

Se observa adem\'{a}s como los estados excitados, ya ocupados por huecos una
vez ocurrida la transici\'{o}n, no resultan degenerados dentro de la
tolerancia de los c\'{a}lculos del programa. Esto estimamos que es debido a que en el marco
en que trabajamos, donde imponemos periodicidad sobre una red puntual de
20x20, el n\'umero de  grados de libertad del sistema resulta finito. Es conocido que para
un n\'{u}mero finito de grados de libertad no se pueden obtener transiciones
de fase discontinuas. Seg\'{u}n resultados de la referencia {[}6{]}, donde los
par\'{a}metros del modelo no estaban a\'{u}n bien especificados, los indicios
de la transici\'{o}n se obtuvieron con mayor aproximaci\'{o}n debido a que
diferencia de energ\'ia de los estados excitados result\'o estar por debajo de la precisi\'{o}n de los c\'{a}lculos
num\'{e}ricos. Esto sugiere que en extensiones futuras del trabajo, al
incrementar el n\'umero de grados de libertad en la soluci\'on del modelo, se precise m\'{a}s
la existencia de una sola soluci\'on cuando el dopaje supere el valor cr\'itico.

\section{Conclusiones}

 En t\'erminos generales, puede concluirse que los resultados obtenidos en este trabajo apoyan  las indicaciones obtendias en \cite{key-6,key-6b} acerca de la existencia  de  una transici\'{o}n de fase cu\'{a}ntica del compuesto $La_{2}CuO_{4}$ controlada por el dopaje con huecos,  en las inmediaciones de $x_{c}=0.2$. Se realiz\'o  una revisi\'{o}n del modelo de una banda introducio en \cite{key-1,key-2,key-3}.
Los par\'{a}metros del modelo fueron reajustados al fijar  el ancho de banda de Matheiss a $3.8$$\,\, eV$, el gap del
estado normal antiferromagn\'{e}tico aislante del $La_{2}CuO_{4}$ a
$2.0\,\,$$eV$ y el valor observado de la constante dil\'{e}ctrica de este material
 a 21.   Se repiti\'o estudio el sistema como funci\'on del  dopaje con huecos realizado en las
 referencias  \cite{key-6,key-6b}, presentando la evoluci\'{o}n de los estados uniparticulares electr\'{o}nicos y de
la superficie de Fermi del sistema en las fases AFA y PPG para un amplio rango
de la concentraci\'{o}n de huecos. Nuevamente, obtuvimos que en la zona de bajo dopaje, los
huecos comienzan a ocupar los estados situados en la frontera de la zona de
Brillouin y que en el estado AFA son los que cargan con el antiferromagnetismo
m\'{a}s intenso. Los estados ocupados en ambas fases tienden gradualmente a hacerse  degenerados cuando el dopaje crece. La energ\'ia total de ambos estados coincide  para una concentraci\'{o}n de huecos cercana a
$x_{c}=0.2$.  Los resultados indican as\'{\i} la presencia de una
transici\'{o}n de fase cu\'{a}ntica del compuesto $La_{2}CuO_{4}$ que pasa de un
estado b\'{a}sico aislante con correlaciones antiferromagn\'{e}ticas a un
estado paramagn\'{e}tico met\'{a}lico en un punto cr\'{\i}tico de
concentaci\'{o}n de huecos  que se encuentra dentro del Domo superconductor.  Puede concluirse
que el presente estudio, en conjunto con el realizado en las referencias \cite{key-6,key-6b},  contribuye  aclarar un importante problema abierto en la F\'isica de los superconductores de alta temperatura: el relativo a la naturaleza y propiedades  del llamado estado de pseudogap  en dichos materiales \cite{key-38}.

\section{Agradecimientos}
   Queremos expresar nuestro agradecimiento a colegas cuyos comentarios han sido de utilidad
durante  el desarrollo de este trabajo. Entre ellos se encuentran:  C. Rodr\'iguez-Castellanos, E. Altshuler,
A. Mart\'inez, Ll. Uranga, V. Mart\'inez, E. Fradkin, J.  Kroha, A. LeClair,  M. D. Coutinho. Tambi\'en se agradece el apoyo recibido para la labor desde el  Network N-35 de la "Office of External Activities" (OEA) del "International Centre for Theoretical Physics" (ICTP), en  Trieste, Italia; as\'i como del Proyecto Nacional de Ciencias Exactas "Teor\'ia Cu\'antica de Campos y de Muchos Cuerpos en Astrof\'isica, F\'isica de Altas Energ\'ias y F\'isica de la Materia Condensada" del Ministerio de Ciencia Tecnolog\'ia y Medio Ambiente (CITMA), La Habana, Cuba.

\end{document}